\documentclass[useAMS,usenatbib,usegraphicx]{mn2e}

\title[Sample variance, source clustering and the counts of faint radio sources]{Sample variance, source clustering and their influence on the counts of faint radio sources}
\author[Heywood, Jarvis \& Condon]{Ian Heywood\thanks{{\tt ianh@astro.ox.ac.uk}}$^{1,2}$, Matt J.~Jarvis$^{1,3}$ and James J.~Condon$^{4}$\\
$^{1}$Astrophysics, Department of Physics, University of Oxford, Keble Road, Oxford, OX1 3RH\\
$^{2}$Department of Physics and Electronics, Rhodes University, PO Box 94, Grahamstown, 6140, South Africa\\
$^{3}$Physics Department, University of the Western Cape, Private Bag X17, Bellville 7535, South Africa\\
$^{4}$National Radio Astronomy Observatory, 520 Edgemont Road, Charlottesville, VA 22903, USA}
\begin{document}

\date{Accepted 2013 February 08. Received 2013 February 07; in original form 2013 January 28}

\pagerange{\pageref{firstpage}--\pageref{lastpage}} \pubyear{2013}

\maketitle

\label{firstpage}

\begin{abstract}

The shape of the curves defined by the counts of radio sources per unit area as a function of their flux density was one of the earliest cosmological probes. Radio source counts continue to be an area of astrophysical interest as they can be used to study the relative populations of galaxy types in the Universe (as well as investigate any cosmological evolution in their respective luminosity functions). They are also a vital consideration for determining how source confusion may limit the depth of a radio interferometer observation, and are essential for characterising the extragalactic foregrounds in Cosmic Microwave Background experiments. There is currently no consensus as to the relative populations of the faintest (sub-mJy) source types, where the counts show a turn-up. Most of the source count data in this regime are gathered from multiple observations that each use a deep, single pointing with an interferometric radio telescope. These independent count measurements exhibit large amounts of scatter (factors of order a few) that significantly exceeds their respective stated uncertainties. In this article we use a simulation of the extragalactic radio continuum emission to assess the level at which sample variance may be the cause of the scatter. We find that the scatter induced by sample variance in the simulated counts decreases towards lower flux density bins as the raw source counts increase. The field-to-field variations make significant contributions to the scatter in the measurements of counts derived from deep observations that consist of a single pointing, and could even be the sole cause at $>$100~$\mu$Jy. We present a method for evaluating the flux density limit that a radio survey must reach in order to reduce the count uncertainty induced by sample variance to a specific value. We also derive a method for correcting Poisson errors on source counts from existing and future deep radio surveys in order to include the uncertainties due to the cosmological clustering of sources. A conclusive empirical constraint on the effect of sample variance at these low luminosities is unlikely to arise until the completion of future large-scale radio surveys with next-generation radio telescopes.

\end{abstract}


\section{Introduction}
\label{sec:intro}

Astrophysical radio emission, at least that which we observe away from the plane of the Milky Way, tends to originate from extragalactic objects at great distances. The differential counts\footnote{i.e.~the number of sources per unit area on the sky with flux densities in the interval $S$~$\rightarrow$~$S+dS$.} of these distant radio sources formed one of the earliest cosmological probes (e.g.~Longair, 1966). In a non-expanding Euclidean universe\footnote{A Euclidean universe filled with sources of luminosity $L$ with number density $n$ contains $N$~=~4$\pi$$n$$d^{3}$/3 such sources out to distance $d$. Since the flux $S$~=~$L$$/$4$\pi$$d^{2}$ it is trivial to show that $N(S)$~$\propto$~$S^{-3/2}$.} populated with non-evolving sources we would see the integrated source counts $n(S)$ scaling with source flux density $S$ according to the relationship $n(S)~\propto~S^{-3/2}$. Observed departures from this relationship thus inform on the geometry of the Universe, and radio source counts were being invoked as early as the 1950s as one of the key evidential cruxes in the Big Bang versus Steady State debate (Ryle \& Clarke, 1961), a cosmological contention that was eventually effectively ended by the discovery of the Cosmic Microwave Background (CMB) radiation (see e.g.~Longair, 2004).

Source counts are thus an area of study that is almost as old as the science of radio astronomy itself. Today the primary interest in source counts (across the whole electromagnetic spectrum) stems from the need to determine the contributions that different galaxy populations make to the total number of objects in the Universe, in particular the relative numbers of star-forming galaxies and those harbouring active-galactic nuclei (AGN), and how the luminosity functions of these populations evolve over cosmic time. Radio source counts are essential for foreground subtraction in CMB experiments, and are also vital for determining where confusion becomes a fundamental limitation in a radio synthesis image. This may occur either due to classical confusion imposed by the sources at the faint end of the distribution that lie within the target field (Condon, 2009), or due to the presence of point spread function sidelobes associated with the brighter sources that lie in distant regions of the array primary beam (Smirnov et al, in prep.). This is particularly relevant at present as we await the arrival of the next-generation of radio instruments. These have been designed to deliver ultra-deep radio imaging and fast survey speeds by virtue of their extreme sensitivities and novel detector technologies, eventually culminating in the deployment of the Square Kilometre Array (SKA; Dewdney et al., 2009).

The faint end of the source count distribution is of particular interest, and there are many publications on the nature of the sub-mJy source population. The 1.4 GHz counts exhibit a turn up at $<$1 mJy (e.g.~Windhorst et al., 1984; Hopkins et al., 1999), that persists at higher frequencies (e.g.~Fomalont et al., 2002; Heywood et al. 2013). Many publications assert the nature of the source population at these levels and it is generally accepted that this is due to the increasing dominance of star-forming galaxies over AGN at these low luminosities (e.g.~Seymour et al., 2008; Padovani et al., 2009), although radio-weak AGN and FR-I type galaxies may make still make significant contributions (Jarvis \& Rawlings, 2004; Simpson et al., 2006; Gendre \& Wall, 2008; Smol{\v c}i{\'c} et al., 2009). There is however no clear consensus as to the relative fractions that these objects occupy.

Additional interest in the faintest end of the radio source counts was recently stimulated due to the balloon-borne Absolute Radiometer for Cosmology, Astrophysics and Diffuse Emission (ARCADE2; Fixsen et al., 2011) experiment which detected a significant excess in the sky brightness temperature at 3~GHz (Seiffert et al., 2011). These data suggest that if the result is genuine there must be a significant population of hitherto unknown faint radio emitters responsible for the excess (Vernstrom et al, 2011). Condon et al.~(2012) performed a probability of deflection ($P(D)$; Scheuer, 1957) analysis of a confusion-limited Very Large Array (VLA) image at 3 GHz with a depth of 1 $\mu$Jy. Their results suggest that if the ARCADE2 result is indeed produced by a population of discrete radio sources then they are exceptionally numerous, not associated with known galaxies and must have 1.4~GHz flux densities of $<$0.03~$\mu$Jy.

Clearly there remains much to learn from surveys of extragalactic radio sources in the $\mu$Jy regime. Examination of the differential source counts from multiple surveys immediately highlights an issue that blights the current data: interpretation of the measured source counts at flux densities $<$1~mJy proves difficult when the derived source counts from survey to survey do not agree to within their respective errors. Possible explanations for the scatter include different calibration accuracies, uncertainties in the method of correcting for the array primary beam and smearing effects (e.g.~Section 2.4, Fomalont et al., 2006), correction of detection thresholds due to resolved sources (e.g.~Section 3.2, Bondi et al., 2008), as well as non-instrumental considerations such as the clustering bias of the sources in the field, i.e.~due to sample variance.

Avoiding sample variance in faint source counts requires a large-area sky survey down to sub-mJy depths, which would require multiple, deep pointings on existing radio telescopes. Condon (2007) investigated the effect of sample variance by measuring the count fluctuations in 17 non-overlapping VLA pointings from the Spitzer First Look Survey and determined that the scatter due to sample variance is (1.07~$\pm$~0.26) times the fluctuations expected in the absence of source clustering, concluding that the field-to-field variations are likely to be non-cosmic in origin.
We take a different approach to quantifying the effect of sample variance by exploiting an existing extragalactic sky simulation in order to present a simple measurement of the scatter induced in the measured counts. For an in-depth review of the subject of radio source counts we refer the reader to de Zotti et al.~(2010).

\section{Method and results}
\label{sec:results}

\begin{figure*}
\begin{center}
\setlength{\unitlength}{1cm}
\begin{picture}(16,7.0)
\put(-1.6,-0.7){\includegraphics{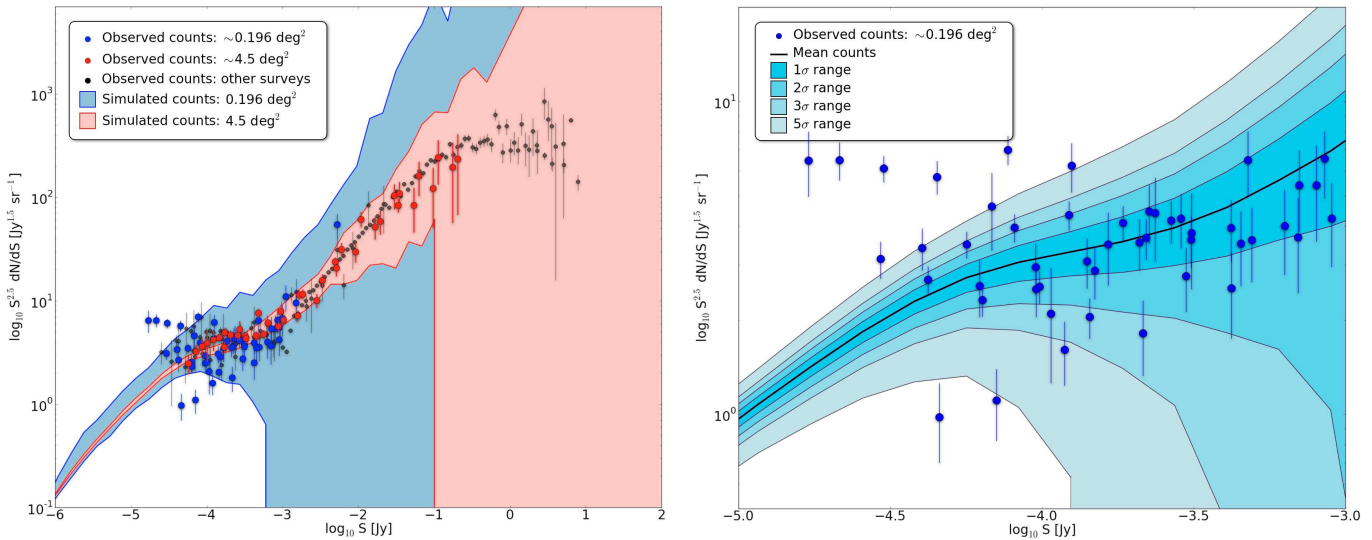}}
\end{picture}
\caption{\label{fig:source_counts} {\bf Left panel:} The data points and the corresponding error bars show the observationally-derived Euclidean-normalized differential source counts at 1.4 GHz from the publications listed in Section \ref{sec:results}. The colours correspond to the three bins into which the observations are divided. The blue points correspond to counts derived from a single VLA pointing, the red points are derived from surveys covering 4 -- 4.5 square degrees, and the black points are from various other (generally larger area) surveys, displayed here in order to present the full source count distribution. The blue and red shaded areas show the full range of source counts derived from independent samplings of a the extragalactic sky simulation of Wilman et al.~(2008; 2010) with areas matched to the blue and red observational data points. {\bf Right panel:} This panel zooms in on the 10~$\mu$Jy -- 1~mJy flux density region. The blue data points are the same as those on the left hand panel. The black line here shows the mean simulated source counts and the shaded regions that surround it correspond to 1, 2, 3 and 5 standard deviations as measured from the 1936 source count measurements in each bin. The data shown on this figure are available from the authors.}
\end{center}
\end{figure*}

We investigate the effect of sample variance on the scatter in the measured source counts by comparing observationally-derived measurements with matched samples drawn from an existing simulation of the extragalactic radio sky.

The data points and associated error bars on Figure \ref{fig:source_counts} show the Euclidean-normalized differential source counts from a variety of radio surveys at 1.4 GHz. The observational source count data that we use for comparison is drawn from fourteen individual studies, most of which are conveniently tabulated by de Zotti et al.~(2010). The solid angle sky coverage of the individual surveys are partitioned into three bins: those that resulted from a single, deep pointing with the VLA, resulting in a nominal survey area of approximately 0.196 deg$^{2}$ (hereafter known as the `deep' bin; Mitchell \& Condon, 1985; Biggs \& Ivison, 2006; Fomalont et al., 2006; Kellermann et al., 2008; Owen \& Morrison, 2008; Seymour et al., 2008; Ibar et al., 2009), surveys covering approximately 4-4.5 deg$^{2}$ (hereafter referred to as the `broad' bin; Ciliegi et al., 1999; Gruppioni et al., 1999; Hopkins et al., 2003), and finally surveys that were in general conducted over sky areas that exceeded the footprint of the simulation described below, and thus cannot be compared. These include the source counts derived from the FIRST survey (White et al., 1997), as well as those from the targeted surveys of Bridle et al.~(1972). Also plotted on the figure are the counts from the 2 deg$^{2}$ radio survey of the COSMOS field (Schinnerer et al., 2004; Bondi et al., 2008).

Counts from surveys in the deep and broad bins are plotted on Figure \ref{fig:source_counts} as the blue and red points respectively. The smaller black points correspond to all other surveys. Immediately apparent from this selection and colour-coding alone is the large spread in source counts derived from the deep sample. This is the issue we aim to address with the simulation.

Our next step is to compare these measured values to matched samples of simulated source counts. For this we make use of the semi-empirical extragalactic sky simulation (hereafter referred to as `the simulation') of Wilman et al.~(2008; 2010)\footnote{The simulation database can be accessed online via {\tt http://s-cubed.physics.ox.ac.uk}.}. Briefly, the simulation uses observed and extrapolated luminosity functions to populate an evolving dark matter skeleton with various galaxy types. The 20~$\times$~20 deg$^{2}$ sky area of the simulation contains $\sim$260 million sources down to a flux density limit of approximately 10 nJy.

We extract multiple sky patches with areas of 0.196 and 4.5 deg$^{2}$ from the simulation for comparison to the measured counts in both the deep and broad bins. This process results in 1936 and 64 unique source catalogues for the deep and broad samples respectively. For each of these simulated source subsets we compute the Euclidean-normalized differential source counts in 58 logarithmically-spaced flux density bins from 10 nJy to 100 Jy. For each bin the maximum and minimum value of the counts delineate a region on the left hand panel of Figure \ref{fig:source_counts} that corresponds to the possible range of field-to-field fluctuations in the source counts of a survey of matched area. This is plotted for both the deep bin (blue area) and the broad bin (red area). 
We stress that this process is not blighted by the biases inherent in deriving accurate counts from observations, such as those mentioned briefly in Section \ref{sec:intro}, and the scatter will be induced purely by the source clustering, itself governed by the underlying model dark matter density field upon which the simulated galaxy population is placed. Our chosen bin widths are well matched to those used in the observations: for every flux density bin used in the set of observations we calculate the ratio of that bin width to that of the simulated bin that is closest to it in terms of central flux, and the mean value of these ratios over all bins considered is 0.96 with a median value of 0.83.

\begin{table*}
\begin{minipage}{\textwidth}
\centering
\caption{Survey depths (detection thresholds, not rms sensitivities) required to restrict the scatter in the source counts imposed by sample variance to values of 1\%, 5\%, 10\% and 25\% of the mean value of the source counts in that flux density bin. These are presented as a function of survey area, and all values are in $\mu$Jy. The values are derived from polynomial fits in log space to the measured ratios of the standard deviation to the mean. Polynomial coefficients are also provided, see Figure \ref{fig:polyfits} and the text for details.\label{tab:areas}}
\begin{tabular}{ccccccccc} \hline
Area (deg$^{2}$) & $S^{1\%}_{limit}$ & $S^{5\%}_{limit}$ & $S^{10\%}_{limit}$ & $S^{25\%}_{limit}$ & $p_{1}$ & $p_{2}$ &$p_{3}$ &$p_{4}$ \\ \hline

0.1 & --    & 2.5    & 17.96    & 155.1    & -0.00842 & -0.07982 & 0.20276 & 0.85972 \\
0.3 & --    & 10.31  & 61.56    & 500.0    & -0.00809 & -0.07383 & 0.21948 & 0.63105 \\
0.5 & --    & 17.96  & 107.2    & 870.6    & -0.00872 & -0.08273 & 0.17371 & 0.4505 \\
1.1 & --    & 45.24  & 253.9    & 2193     & -0.00962 & -0.09393 & 0.12197 & 0.20776 \\
1.5 & 0.102 & 69.63  & 367.5    & 3174     & -0.01073 & -0.10866 & 0.0578 & 0.05038 \\
2.1 & 0.348 & 94.75  & 531.8    & 5197     & -0.0095  & -0.09286 & 0.10834 & 0.0151 \\
3.1 & 1.055 & 155.1  & 818.5    & 7521     & -0.0102  & -0.09855 & 0.10431 & -0.03349 \\
4.1 & 3.008 & 211.1  & 1113     & 13930    & -0.01431 & -0.15594 & -0.13708 & -0.41124 \\
4.9 & 3.848 & 270.1  & 1425     & 16750    & -0.01321 & -0.13809 & -0.05466 & -0.34036 \\

\hline
\end{tabular}
\end{minipage}
\end{table*}

The right hand panel of Figure \ref{fig:source_counts} shows a zoomed-in region covering the 10~$\mu$Jy to 1~mJy flux density region. Again the blue points show the observed source counts for single pointing experiments. The mean value of the simulated counts from the 1936 independent distributions in each bin is shown by the black line. The shaded regions surrounding this correspond to 1, 2, 3 and 5 times the standard deviation of the count measurements. 

\begin{figure}
\nonumber
\centering
\includegraphics[width= \columnwidth]{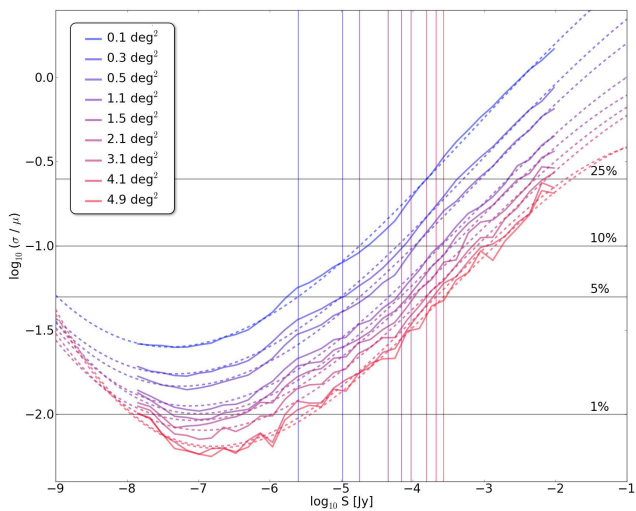} 
\caption{\label{fig:polyfits}The solid lines show standard deviations ($\sigma$) per flux density bin for a range of theoretical (colour coded) survey areas, expressed as a fraction of the mean source counts ($\mu$) in that bin. This plot essentially shows the detection threshold that a survey needs to reach to limit the uncertainty induced in the counts by sample variance to a specific level. Polynomials are fitted to the base-10 logarithms of the distributions, as shown by the dashed lines. Details are given in the text and coefficients are provided in Table \ref{tab:areas}, along with depth requirements for 1, 5, 10 and 25\% values of the count uncertainty (delineated by the horizontal lines). The vertical lines show the detection thresholds that must be reached in order to deliver 5\% uncertainty for each area.}  
\end{figure}

Figure \ref{fig:source_counts} clearly shows that the scatter induced in the source counts by the clustering of radio sources across the sky for a survey of fixed area is thus strongly dependent on the depth of the survey, due to the unmodified surface density of radio sources rising with decreasing flux density. Observational challenges notwithstanding, larger areas are required to accurately quantify the counts of faint radio sources. Count fluctuations induced by sample variance are significant enough to dominate the observed scatter at flux densities above $\sim$100~$\mu$Jy, and contribute significantly below this. Notable outliers on Figure \ref{fig:source_counts} are the anomalously high and rising count values from Owen \& Morrison (2008). The $P(D)$ analysis of Condon et al.~(2012) was conducted over the same field as the Owen \& Morrison (2008) observations, partially motivated by the prospect of confirming the high counts previously seen in that region. Condon et al.~(2012) determine new counts with their 8" resolution VLA C-array observations that are a factor of $\sim$4 lower than the existing ones derived from the multi-configuration, 1.6" resolution observations of Owen \& Morrison (2008), and speculate that overestimation of the resolution corrections are responsible for the discrepancy. 

There is a deviation of the measured broad area counts from the corresponding simulated samples in the left hand panel of Figure \ref{fig:source_counts} below approximately 150~$\mu$Jy. At this depth the broad area counts are drawn solely from the Phoenix Deep Survey (Hopkins et al., 2003). This survey includes a deeper tier that has an effective area that is notably less ($\sim$1--1.5 deg$^{2}$) than the 4.5 deg$^{2}$ probed by the multiple samples of the simulation, and it is from this smaller, deeper region that these counts originate. The deviation illustrates that even on scales of $\sim$1 deg$^{2}$ the sampling variation in the counts is not negligible.

As noted by Wilman et al.~(Section 4, 2008), in order to predict the behaviour of the radio sky at levels that are beyond present observation requires extrapolation of the known luminosity functions. We naturally cannot rule out departures of the simulation from reality below the limits of the observationally measured source counts. Our results are also sensitive to the accuracy of the clustering model in the simulation. Wilman et al.~(2008) test the validity of the source clustering by comparing the simulated and measured angular two point correlation functions, and find good agreement. For further details, including potential (less significant) limitations of the simulation we refer the reader to Wilman et al.~(2008).   

Note also that the brightest end of the source counts also have uncertainties in the measurements comparable to those associated with the faintest counts. The effect that causes the large scatter is analogous at both ends of the scale: in the case of the bright sources it is a combination of small effective survey volumes for nearby sources and the intrinsic rarity of extremely bright sources at large distances, resulting in low number counts in both scenarios.

The following two subsections broaden the utility of the above results by presenting a pair of tools for observers who wish to carry out deep radio surveys in order to investigate the faint radio source population.

\subsection{Optimisation of survey area according to flux density detection threshold}
\label{sec:optimise}

Here we present a method for approximately evaluating the area that a survey of a given detection threshold must cover in order to limit the uncertainty in the counts induced by sample variance to a certain level. The standard deviation derived from the multiple count samples per flux density bin ($\sigma$) is expressed as a fraction of the mean count value ($\mu$) in that bin, and these data are plotted in log space as solid lines on Figure \ref{fig:polyfits}. These calculations are performed for a representative group of nine survey areas ranging from 0.1 to 4.9 deg$^{2}$, as listed on the legend of Figure \ref{fig:polyfits}. Testing sky areas larger than this becomes problematic as the number of independent catalogues that can be extracted from the simulation decreases with sky area. This is reflected in the increasing ripple levels of the curves on Figure \ref{fig:polyfits} as the sky area increases.

A good approximation to the measured curves is provided by a least-squares fitted polynomial of the form
\begin{equation}	
\mathrm{log}(\mu/\sigma)~=~p_{1} + p_{2}\mathrm{log}(S) + p_{3}\mathrm{log}(S)^{2} + p_{4}\mathrm{log}(S)^{3}.
\end{equation}
The fitted curves are shown by the dashed lines on Figure \ref{fig:polyfits}. The coefficients $p_{n}$ are provided in Table \ref{tab:areas} for the nine survey areas, allowing the approximate uncertainties to be calculated for arbitrary surveys. As this is a polynomial fit it should not be used to extrapolate outside the range of the data to which it was fitted, however the lower limit of 10 nJy is the formal flux-density limit of the simulation, and the source counts are generally well constrained observationally beyond the 10 mJy upper limit and up to the rare $>$1~Jy population.

Table \ref{tab:areas} also lists the survey limits required to reduce the scatter in the source counts to 1, 5, 10 and 25\% of the mean values (shown by the horizontal lines on Figure \ref{fig:polyfits}) for the nine hypothetical surveys. To illustrate how these limits are determined the 5\% case is presented as an example by the colour coded vertical lines on Figure \ref{fig:polyfits}. Note that the four smallest sky areas do not provide the accuracy to ever reach a 1\% uncertainty within the limits of the simulation, hence the missing values in Table \ref{tab:areas}.

\subsection{Corrections for Poisson uncertainties in order to include the effects of source clustering}
\label{sec:sigma_CL}

The sample variance is equivalent to the variance of the counts in the cells into which the simulation is divided, and consists of two components, namely the Poisson variance and a second contribution caused purely by the cosmological clustering of the sources. In this section we provide an estimate of the contribution to the sample variance that is solely due to source clustering as a function of flux density and survey solid angle. This allows existing and future experiments that measure the counts of faint radio sources to correct their Poisson errors in order to include clustering effects.

The 1$\sigma$ percentage errors due to both Poisson scatter ($\sigma^{\%}_{P}$) and sample variance ($\sigma^{\%}_{S}$) can be calculated for the simulated Euclidean-normalized differential source counts for each flux density bin. An estimate of pure Poisson errors that does not include the effects of source clustering is derived by randomising the position of each source in the simulation and measuring the variance of the counts in each cell. This procedure is carried out 100 times and the mean variance is used to calculate the 1$\sigma$ Poisson percentage error $\sigma^{\%}_{P}$. The sample variance uncertainty is taken as the standard deviation of the individual count values in each cell of the unperturbed simulation, as per the 1$\sigma$ limits presented in Section \ref{sec:optimise}. These calculations are performed in flux density bins with a logarithmic width of 0.2 Jy over the full flux-density range of the simulation. 

How can the contribution to the sample variance that is purely due to cosmological source clustering be distilled? We assume that the source clustering multiplies the number of galaxies in each independent cell by a factor $f$ that has a mean value of 1. The rms percentage scatter in this factor is denoted by $\sigma^{\%}_{CL}$, and is independent of the raw source counts in any given bin (and thus independent of the Poisson errors). Furthermore, the factor $f$ is assumed to be a function of flux density that varies slowly enough such that $f$ can be treated as constant across each flux density bin in which sources are counted.

\begin{figure}
\nonumber
\centering
\includegraphics[width= \columnwidth]{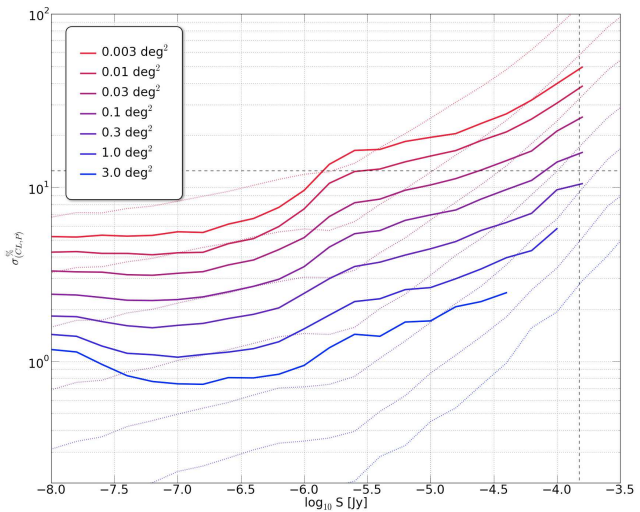} 
\caption{\label{fig:sigma_CL}Values of $\sigma^{\%}_{CL}$ for seven survey solid angles in the range 0.003 to 3.0 deg$^{2}$. The values of $\sigma^{\%}_{CL}$ are for use in Equation \ref{eq:sigma_CL} in order to apply a correction to observationally-derived Poisson errors in order to include the cosmological clustering of sources. The sky areas covered by this plot should ensure that it remains useful for single-pointing observations with future radio telescopes such as MeerKAT and the dish component of the SKA. The faint dotted curves show the mean fractional percentage Poisson errors for comparison to existing theory at the end of Section \ref{sec:sigma_CL}.}  
\end{figure}

If the distribution of radio sources were devoid of any clustering then the Poisson variance ($\sigma^{\%}_{P}$) would be the sole cause of the scatter in the Euclidean-normalized counts ($N_{bin}$) in any given flux-density bin. We assume that the source clustering contributes to the measured variance ($\sigma^{\%}_{S}$) from the simulation in a way that conforms to the behaviour of the $f$ parameter described above, i.e.~the clustering adjusts the measured counts to a value of $f$~$\times$~$N_{bin}$. The sample variance (i.e.~the variance of the counts in each cell of the simulation, $\sigma^{\%}_{S}$) is thus the quadratic sum of the Poisson variance ($\sigma^{\%}_{P}$) and the additional variance due to cosmological clustering ($\sigma^{\%}_{CL}$). It does not drop to zero even in the absence of any source clustering. We can extract the rms percentage scatter in $f$ using error propagation rules:
\begin{equation}
\sigma^{\%}_{CL}~=~\sqrt{(\sigma^{\%}_{S})^{2}~-~(\sigma^{\%}_{P})^{2}}
\end{equation}
since in the absence of clustering the Poisson variance is the sole contributor to the sample variance. The parameter $\sigma_{CL}$ is independent of the choice of bin width, and its values derived from our simulation can be used in conjunction with an observationally-derived value of $\sigma_{P}^{\%}$ to determine
\begin{equation}
\sigma^{\%obs}_{S}~=~\sqrt{(\sigma^{\%obs}_{P})^{2}~+~(\sigma^{\%}_{CL})^{2}},
\end{equation}
i.e.
\begin{equation}
\label{eq:sigma_CL}
\sigma^{\%obs}_{S}~=~\sqrt{\frac{100^{2}}{N^{obs}_{bin}}~+~(\sigma^{\%}_{CL})^{2}}
\end{equation}
where $N^{obs}_{bin}$ is the number of sources in that flux density bin. 

Figure \ref{fig:sigma_CL} shows the values of $\sigma_{CL}^{\%}$ derived from the simulation that are applicable to faint flux density bins (10.0~nJy~$<$~$S_{centre}$~$<$~0.3~mJy) for a range of effective survey solid angles. For a given measurement of the Euclidean-normalized differential source counts, Figure \ref{fig:sigma_CL} can be used in conjunction with Equation \ref{eq:sigma_CL} in order to correct the percentage error estimate ($\sigma^{\%obs}_{S}$) in the observed counts ($N_{bin}^{obs}$) to include clustering effects. We impose the condition that for the derived value of $\sigma_{CL}^{\%}$ to be trustworthy, it must exceed 5$\sigma_{P}^{\%}$. This is to account for the fact that the Poisson errors derived from flux density bins containing average counts of $<$1 cannot be reliably used. These conditions lead to the cut-offs in the lines on Figure \ref{fig:sigma_CL}. The cut-offs manifest themselves at fainter flux densities with smaller survey solid angles as the raw source counts per bin decrease with sky coverage.

The seven sky survey areas in Figure \ref{fig:sigma_CL} cover the range 0.003 to 3.0 deg$^{2}$. The smallest areas are chosen to make the figure relevant for the current deepest observations, where the faintest sources are detected in effective areas much smaller than the primary beam size. The broader areas make the plot relevant for future radio continuum surveys with MeerKAT (13.5~m dishes) and the SKA (15~m dishes)\footnote{The Australian SKA Pathfinder (ASKAP) is a special case as it has been designed to deliver an instantaneous field of view at 1.4 GHz of $\sim$30 deg$^{2}$. The sample variance contribution due to the clustering of cosmological sources is not likely to be an issue for the surveys that are planned for it.}. 

We can compare our predictions for the effects of source clustering to the measurement of Condon (2007). Seventeen independent pointings of approximately 0.2 deg$^{2}$ each were extracted from the Spitzer First Look Survey, and with approximately 100 sources per field with a flux density limit of 150 $\mu$Jy, our simulation predicts a $\sigma_{CL}$ value of approximately 12.5\%, as shown by the intersecting dashed lines on Figure \ref{fig:sigma_CL}. Applying these values to Equation \ref{eq:sigma_CL} results in a percentage error in the observed counts of $\sigma_{S}^{\%obs}$~=~16\%. This is slightly higher than but still broadly consistent with the observed value of (10.7~$\pm$~2.6)\%.

The shapes of the $\sigma^{\%}_{CL}$ curves on Figure \ref{fig:sigma_CL} are worthy of comment as they say something about the clustering strength of radio sources as a function of their flux densities. The plot shows the area-dependent trend that one would instinctively expect. The effect of source clustering rises with flux density although this is not a smooth change over the plotted range. This is likely due to the brighter end of the source counts likely being dominated by more massive elliptical galaxies that are more strongly clustered than the faint sources, the less clustered star-forming spiral galaxies. 

Finally, we compare the trend that these lines exhibit to existing theory. Clustering will increase the variance of the source counts in each individual cell. If each cell contains $N$ sources in a solid angle $\Omega$ and a (fairly narrow) flux-density range $\Delta S$, then the mean number of sources per cell is
\begin{equation}
 \bar{N} = n(S) \Delta S\, \Omega~.
\end{equation}
The sample variance can be written as the sum of the Poisson variance $\bar{N}$ and the variance caused solely by clustering. Peebles (1980) expresses this in terms of $w(\theta)$, the two-point correlation as a function of angular separation $\theta$:
\begin{equation}
\langle (N - \bar{N})^2 \rangle = \bar{N}
+ {\bar{N}^2 \over \Omega^2} \int w(\theta) d \Omega_1 d \Omega_2~.
\end{equation}
The function $w(\theta)$ is usually approximated by a power-law of the form
\begin{equation}
w(\theta) = A \biggl({ \theta \over {\rm deg}}\biggr)^{-\alpha}~.
\end{equation}
Blake \& Wall (2002a,b) measured $w(\theta)$ in the range $0.1 < \theta {\rm \,(deg)} < 10$ for NRAO VLA Sky Survey (NVSS; Condon et al., 1998) sources stronger than about 10~mJy at 1.4~GHz and found $A \approx 1.0 \times 10^{-3}$, $\alpha \approx 0.8$. \citet{bla04} combined Sydney University Molonglo Sky Survey (SUMSS; Bock, Large \& Sadler, 1999), NVSS, and Westerbork Northern Sky Survey (WENSS; Rengelink et al., 1997) data to estimate a slightly larger $A \approx 1.6 \times 10^{-3}$ and
a slightly steeper $\alpha \approx 1.1$.

Following de Zotti et al.~(2010), we note that the fractional variance
\begin{equation}
{\langle(N - \bar{N})^2\rangle \over \bar{N}^2} = 
{1 \over \bar{N}} + {1 \over \Omega^2} \int w(\theta) d \Omega_1 d \Omega_2
\end{equation}
has the advantage that the clustering term does not explicitly depend on $\bar{N}$ or $\Delta S$. Using our notation
\begin{equation}
{\langle(N - \bar{N})^2\rangle \over \bar{N}^2} = 
{1 \over \bar{N}} + \sigma_{CL}^2~,
\end{equation}
where
\begin{equation}
\sigma_{{CL}}^2 =  {1 \over \Omega^2} \int w(\theta) d \Omega_1 d \Omega_2
\approx 2.36 A \biggl({\Omega \over {\rm deg^2}}\biggr)^{-\alpha/2}
\end{equation}
is the fractional variance contributed by clustering alone. Thus
\begin{equation}
\sigma_{CL}^{\%} \approx 5 \biggl({\Omega \over 
{\rm deg^2}}\biggr)^{-\alpha/4}
\end{equation}	
declines more slowly with $\Omega$ than the Poisson scatter, which
is proportional to $\Omega^{-1/2}$, as is reflected in our results in Figure \ref{fig:sigma_CL}, in which the fainter dotted lines show the mean fractional percentage Poisson errors.

\section{Concluding remarks}

Observationally-derived values of the counts of faint radio sources exhibit levels of scatter that can be up to a factor of several greater than the quoted uncertainties in the counts. We have provided an estimate of the scatter induced in the counts of faint radio sources due to the sample variance induced by cosmological source clustering by using many independent samples of an extragalactic sky simulation, and comparing these results to matched observations. The deepest observations to date have been carried out using single deep pointings with the VLA. The fluctuations induced by sample variance in the counts derived from such an observation may be large enough to completely explain the observed scatter at flux densities above approximately 100~$\mu$Jy, and we have quantified their contribution as a function of survey area below this level.

We have presented a method for estimating the count uncertainty induced by sample variance for an arbitrary radio survey, or reciprocally, for determining the depth that a radio survey of fixed solid angle coverage must reach in order to limit the count uncertainty. We have also derived a method for correcting Poisson errors in order to include the effects of source clustering. This method is applicable to the deepest surveys that exist today and should remain applicable for future deep continuum surveys with the VLA, MeerKAT and the SKA, down to survey flux density limits of 0.1~$\mu$Jy. We stress again the distinction between survey flux density limits and the rms sensitivity of the corresponding radio images when applying these methods.

The amount that cosmological clustering affects the counts is as one would expect strongly dependent on survey area but also on flux density limits, likely due to the preferential clustering of massive elliptical galaxies at the brighter end, with the less clustered star-forming spiral galaxies dominating the fainter counts. 

The method for correcting Poisson uncertainties is broadly consistent with the observationally-derived measurement of the count fluctuations presented by Condon (2007), who concluded that human-induced  instrumental calibration and interpretative differences are likely to dominate the scatter. Such effects are certainly contributing factors to the difference in published counts in cells between different authors; the potential overestimation of the resolution correction resulting in the very high counts of Owen \& Morrison (2008) being a prime example that is not explained by our results. The sample variance in the case of the deepest surveys such as this is only marginally larger than the actual Poisson variance due to the source counts per bin being very low, counted over effective areas much smaller than the primary beam size.

Current facilities are not suited to deriving a low-uncertainty measurement of the faint radio source counts without an unfeasibly large investment of telescope time. It is likely that the issue will lack an empirical resolution until the completion of the next-generation of legacy radio surveys with future instruments such as ASKAP, MeerKAT and eventually the SKA.

\section*{Acknowledgments}

We thank Andrew Hopkins for refereeing this paper, and suggesting numerous significant improvements to its content. I.H.~thanks the South-East Physics Network (SEPnet) and the University of Oxford. This research has made use of NASA's Astrophysics Data System.


\label{lastpage}

\end{document}